\documentclass[a4paper,11pt]{article}
\pdfoutput=1 

\usepackage{jinstpub} 
\usepackage{titlesec}                     
\usepackage{parskip}
\usepackage{float}
\usepackage{bold-extra}

\usepackage{times, txfonts}
\titlespacing{\section}{0pt}{\parskip}{2pt}
\titlespacing{\subsection}{0pt}{\parskip}{0pt}
\titlespacing{\subsubsection}{0pt}{\parskip}{-\parskip}

\author[a,1]{M. Gul,\note{Corresponding author.}} 
\author[n,1]{G. Gonzalez Blanco,\note{Corresponding author.}}

\author[a]{A. Cimmino,}
\author[a]{S. Crucy,}
\author[a]{A. Fagot,}
\author[a]{A. A. O. Rios,}
\author[a]{M. Tytgat,}
\author[a]{N. Zaganidis,}
\affiliation[a]{Ghent University, Department of Physics and Astronomy, Proeftuinstraat 86, 9000 Ghent, Belgium}

\author[b]{S. Aly,}
\author[b]{Y. Assran,}
\author[b]{A. Radi,}
\author[b]{A. Sayed,}
\affiliation[b]{Egyptian Network for High Energy Physics, Academy of Scientific Research and Technology, 101 Kasr El-Einy St. Cairo Egypt}

\author[c]{G. Singh,}
\affiliation[c]{Chulalongkorn University, Department of Physics, Faculty of Science, Payathai Road, Phatumwan, Bangkok, THAILAND}

\author[d]{M. Abbrescia,}
\author[d]{G. Iaselli,}
\author[d]{M. Maggi,}
\author[d]{G. Pugliese,}
\author[d]{P. Verwilligen,}
\affiliation[d]{INFN, Sezione di Bari, Via Orabona 4, IT-70126 Bari, Italy}

\author[e]{W. V. Doninck,}
\affiliation[e]{Vrije Universiteit Brussel, Boulevard de la Plaine 2, 1050 Ixelles, Belgium}

\author[f]{S. Colafranceschi,}
\author[f]{A. Sharma,}
\affiliation[f]{Physics Department CERN, CH-1211 Geneva 23, Switzerland}

\author[g]{L. Benussi,}
\author[g]{S. Bianco,}
\author[g]{D. Piccolo,}
\author[g]{F. Primavera,}
\affiliation[g]{INFN, Laboratori Nazionali di Frascati (LNF), Via Enrico Fermi 40, IT-00044 Frascati, Italy}

\author[h]{V. Bhatnagar,}
\author[h]{R. Kumari,}
\author[h]{A. Mehta,}
\author[h]{J. Singh,}
\affiliation[h]{Department of Physics, Panjab University, Chandigarh Mandir 160 014, India}

\author[i]{A. Ahmad,}
\author[i]{M. I. Asghar,}
\author[i]{S. Muhammad,}
\author[i]{I. M. Awan,}
\author[i]{H. R. Hoorani,}
\author[i]{W. Ahmed,}
\author[i]{H. Shahzad,}
\author[i]{M. A. Shah,}
\affiliation[i]{National Centre for Physics, Quaid-i-Azam University, Islamabad, Pakistan}

\author[j]{S. W. Cho,}
\author[j]{S. Y. Choi,}
\author[j]{B. Hong,}
\author[j]{M. H. Kang,}
\author[j]{K. S. Lee,}
\author[j]{J. H. Lim,}
\author[j]{S. K. Park,}
\affiliation[j]{Korea University, Department of Physics, 145 Anam-ro, Seongbuk-gu,Seoul 02841, Republic of Korea}

\author[k]{M. S. Kim,}
\affiliation[k]{Kyungpook National University, 80 Daehak-ro, Buk-gu, Daegu 41566, Republic of Korea}

\author[l]{M. Goutzvitz,}
\author[l]{G. Grenier,}
\author[l]{F. Lagarde,}
\author[l]{F. Lagarde,}
\affiliation[l]{Universite de Lyon, Universite Claude Bernard Lyon 1, CNRS-IN2P3, Institut de Physique Nucleaire de Lyon, Villeurbanne, France}

\author[m]{C. U. Estrada,}
\author[m]{I. Pedraza,}
\author[m]{C. B. Severiano,}
\affiliation[m]{Benemerita Universidad Autonoma de Puebla, Puebla, Mexico}

\author[n]{S. Carrillo Moreno,}
\author[n]{F. Vazquez Valencia,}
\affiliation[n]{Universidad Iberoamericana, Mexico City, Mexico}

\author[o]{L. M. Pant,}
\affiliation[o]{Nuclear Physics Division Bhabha Atomic Research Centre Mumbai 400 085, INDIA}

\author[p]{S. Buontempo,}
\author[p]{N. Cavallo,}
\author[p]{M. Esposito,}
\author[p]{F. Fabozzi,}
\author[p]{G. Lanza,}
\author[p]{L. Lista,}
\author[p]{S. Meola,}
\author[p]{M. Merola,}
\author[p]{I. Orso,}
\author[p]{P. Paolucci,}
\author[p]{F. Thyssen,}
\affiliation[p]{INFN, Sezione di Napoli, Complesso Univ. Monte S. Angelo, Via Cintia, IT-80126 Napoli, Italy} 

\author[q]{A. Braghieri,}
\author[q]{A. Magnani,}
\author[q]{P. Montagna,}
\author[q]{C. Riccardi,}
\author[q]{P. Salvini,}
\author[q]{I. Vai,}
\author[q]{P. Vitulo,}
\affiliation[q]{INFN, Sezione di Pavia, Via Bassi 6, IT-Pavia, Italy} 

\author[r]{Y. Ban,}
\author[r]{S. J. Qian,}
\affiliation[r]{School of Physics, Peking University, Beijing 100871, China} 

\author[s]{M. Choi,}
\affiliation[s]{University of Seoul, 163 Seoulsiripdae-ro, Dongdaemun-gu, Seoul, Republic of Korea}

\author[t]{Y. Choi,}
\author[t]{J. Goh,}
\author[t]{D. Kim,}
\affiliation[t]{Sungkyunkwan University, 2066 Seobu-ro, Jangan-gu, Suwon-si, Gyeonggi-do, Republic of Korea}

\author[u]{A. Aleksandrov,}
\author[u]{R. Hadjiiska,}
\author[u]{P. Iaydjiev,}
\author[u]{M. Rodozov,}
\author[u]{S. Stoykova,}
\author[u]{G. Sultanov,}
\author[u]{M. Vutova,}
\affiliation[u]{Bulgarian Academy of Sciences, Inst. for Nucl. Res. and Nucl. Energy, Tzarigradsko shaussee Boulevard 72, BG-1784 Sofia, Bulgaria}

\author[v]{A. Dimitrov,}
\author[v]{L. Litov,}
\author[v]{B. Pavlov,}
\author[v]{P. Petkov,}
\affiliation[v]{Faculty of Physics, University of Sofia,5, James Bourchier Boulevard, BG-1164 Sofia, Bulgaria}

\author[w]{D. Lomidze,}
\author[w]{I. Bagaturia,}
\affiliation[w]{Tbilisi University, 1 Ilia Chavchavadze Ave, Tbilisi 0179, Georgia}

\author[x]{C. Avila,}
\author[x]{A. Cabrera,}
\author[x]{J. C. Sanabria,}
\affiliation[x]{Universidad de Los Andes, Apartado Aereo 4976, Carrera 1E, no. 18A 10, CO-Bogota, Colombia}

\author[y]{I. Crotty,}
\affiliation[y]{Dept. of Physics, Wisconsin University, Madison, WI 53706, United States}

\author[z]{and J. Vaitkus}
\affiliation[z]{Vilnius University, Vilnius, Lithuania}

\title{\boldmath Detector Control System and Efficiency Performance for CMS RPC at GIF++}

\emailAdd{muhammad.gul@cern.ch}

\abstract{In the framework of the High Luminosity LHC upgrade program, the CMS muon group built several different RPC prototypes that are now under test at the new CERN Gamma Irradiation Facility (GIF++). A dedicated Detector Control System (DCS) has been developed using the WinCC-OA tool to control and monitor these prototype detectors and to store the measured parameters data. Preliminary efficiency studies that set the base performance measurements of CMS RPC for starting aging studies are also presented.}

\collaboration[c]{on behalf of the CMS Collaboration}
\proceeding{13$^{\text{th}}$ Workshop on Resistive Plate Chambers and Related Detectors\\
  February 22-26, 2016\\
  Ghent University, Belgium}

\begin{document}
\maketitle
\flushbottom

\section{Introduction}
\label{sec:intro}
\begin{figure}[H]
\centering
\hspace{-0.5cm}
\includegraphics[scale=0.35]{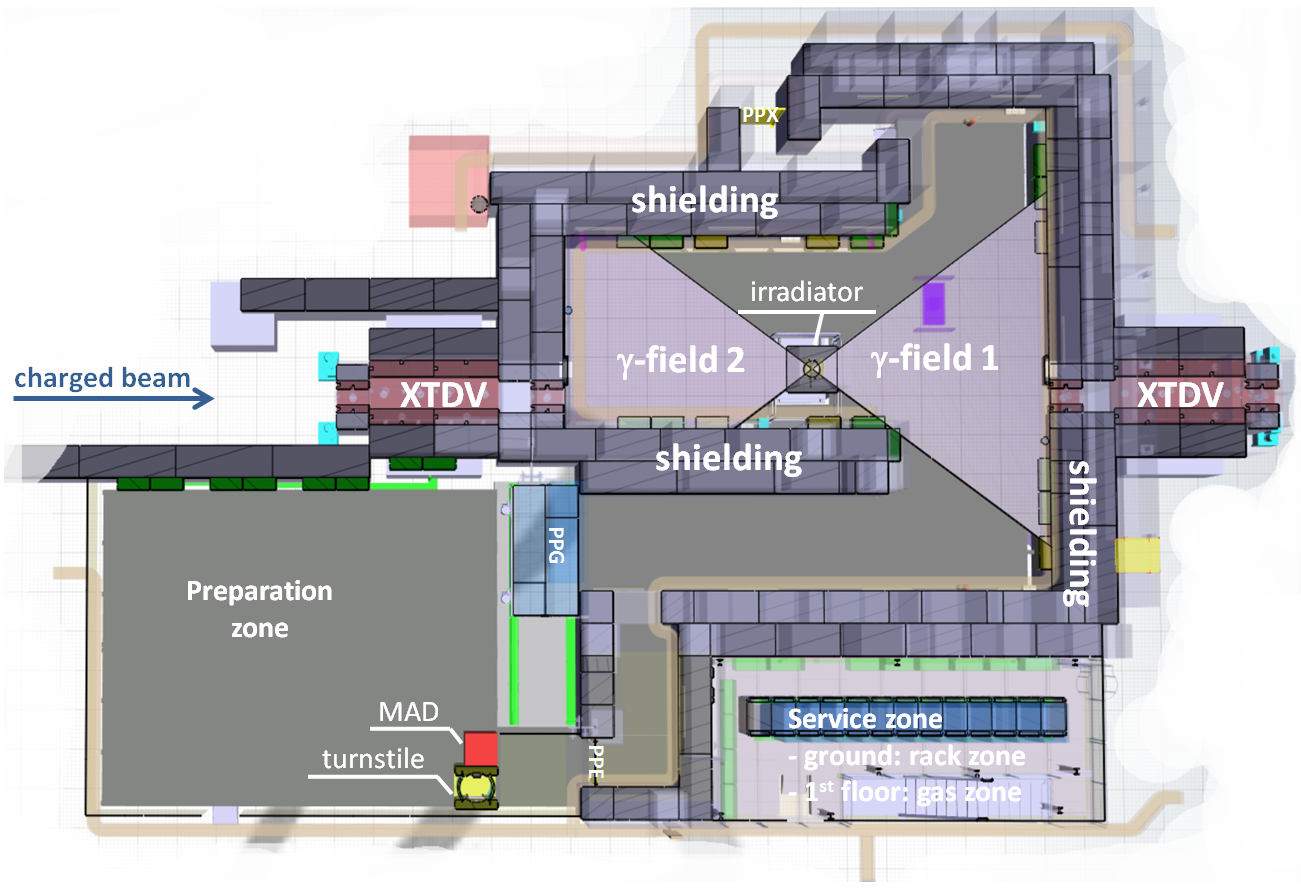}
 \caption{An overview of the GIF++.}
\label{gif}
\end{figure}

The High Luminosity Large Hadron Collider (HL-LHC) machine will induce higher background radiations compared to the current operating conditions. It is important to study the performance and stability of the currently installed and future detectors in a high radiation environment. Focused on these requirements, the CERN Engineering- (EN) and Physics- Department (PH) made a joint project, the Gamma Irradiation Facility (GIF++) \cite{gif}. GIF++ is the new CERN irradiation facility located in the North Area of the CERN Super Proton Synchrotron (SPS). It is a unique place where high energy ($\sim$100 GeV) charged particles (mainly muons) are combined with a high flux of gamma radiation (662 keV) produced by 13.9 TBq $^{137}$Cs source \cite{atlas-gif}. An attenuator system is installed to vary the gamma flux on the two sides of the source independently. A schematic overview of the GIF++ is shown in figure \ref{gif}.\\
The Compact Muon solenoid (CMS) is one of the two general-purpose detectors at the LHC \cite{cms-lhc}, which uses Resistive Plate Chambers (RPCs) along with other detectors \cite{muon-sys} for the muon detection. In April 2015, the RPC detectors were installed at the GIF++ to study the performance at a radiation dose equivalent at 3000 fb$^{-1}$ of the CMS. A dedicated control system has been built to control these detectors and archive the relevant parameters using the WinCC-OA (PVSS) Supervisory Control And Data Acquisition (SCADA) system \cite{wincc-oa}. The system controls high voltage and low voltage supplies and monitors temperature, pressure and humidity of both the RPC gas and the environment. The source status and attenuator values are accessed through the Data Interchange Protocol (DIP), published centrally by the Engineering Department. The RPC gas supply is controlled and monitored by an external WinCC-OA project, that shares relevant parameters with this project. All relevant parameters are archived in a Structured Query Language (SQL) database (DB) for offline analysis.\\
One of the features of the GIF++ RPC DCS system is accessing the source status and attenuator values. Based on this information, RPC performance parameters (efficiency, working point, cluster size and resistivity) are measured. To retrieve the data from the database, a specific algorithm has been developed to synchronize the detector parameters (current and voltage) with the external parameters (temperature, pressure and humidity). It enables to monitor precisely the effect of external parameters on the detector.
 
\section{Motivation}
At HL-LHC, RPCs and associated electronics will operate at much higher luminosity ($\sim$ 5 $\times$ 10$^{34}$ cm$^{-2}$ s$^{-1}$) while accumulating large radiation dose expected from more than 3000 fb$^{-1}$. A precise understanding of the ageing of detector's material and currently installed electronics is needed in intensively high radiation. 
The GIF++ gives the same conditions which will be present at CMS during HL-LHC. The RPCs are irradiated with photons, produced by a $^{137}$Cs radiation source of 13.9 TBq.  At the same time, a high-momentum-particles beam extracted from the SPS, mainly consist of muons, is used to study the performance of the detectors by measuring and comparing their efficiency. The two main gas components of the RPC detector are C$_{2}$H$_{2}$F$_{4}$ and SF$_{6}$ which are going to be gradually banned due to greenhouse effects. An R\&D of different gases mixture is under study at GIF++ to replace the two gases by eco-friendly gases. When a good candidate for a new RPC gas mixture is identified, the study will be further carried out at GIF++ to include long-term assessment of the radiation tolerance of the chambers operated with the new gas mixture. Further details can be found in \cite{phase2}.

\section{WinCC-OA}
Large experiments at CERN use WinCC-OA as a tool to develop control systems. It allows for the description of a device in terms of a data point, with data point elements representing its parameters. These can then be addressed directly to write to and read from the corresponding device. Parameters of interest can then be stored by WinCC-OA in an internal database for offline analyses. WinCC-OA provides the facility to build a graphical user interface (GUI).

\section{The CMS RPC DCS project at GIF++}
\begin{figure}[H]
\centering
\hspace{-0.5cm}
\includegraphics[scale=0.4,trim=60 30 60 30,clip]{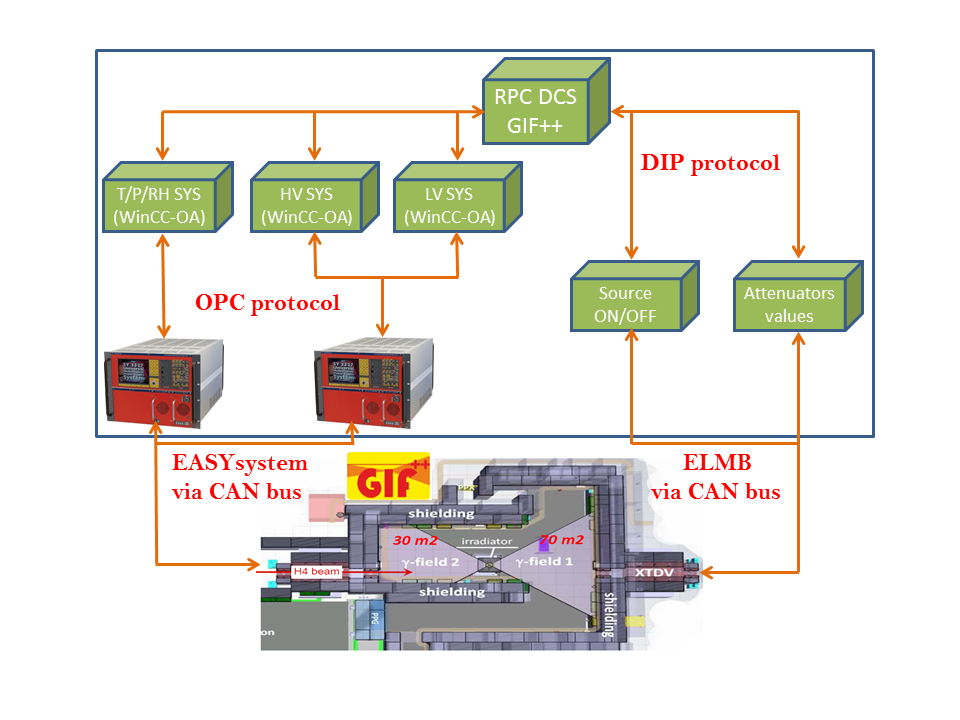}\\
 \caption{DCS project overview.}
\label{DCS_sys}
\end{figure}

The CMS RPC DCS at GIF++ has been developed by using WinCC-OA 3.11 and extended with the standard Joint Control Project (JCOP) framework \cite{jcop}. The JCOP framework provides extra functions such as standardized Finite State Machine (FSM), the additional Graphical User Interface (GUI), the alarm handlers and the ORACLE database interface \cite{g-polese}. The high voltage and low voltage power supplies used in the RPC GIF++ setup consist of CAEN SY1527 mainframe modules as well as CAEN EASY modules, with additional ADC modules used to read out gas and environmental sensors (pressure, temperature and humidity). The project has access to the hardware registered through an Object Linking and Embedding (OLE) for Process Control (OPC) server provided by CAEN using the OPC protocol \cite{opc}. The project controls the high voltage and low voltage system through the OPC protocol. The environmental and gas sensors ( for pressure, temperature and humidity) are also readout via the OPC protocol. The source status and attenuator values are available centrally via the Data Interchange Protocol (DIP). The project has been designed as a distributed one in order to be able to communicate with other projects and to read valuable information. Communication has been established with the central GIF++ DCS, such that the information from the gas system, like flow rates are readable.\\
The Finite State machine (FSM) hierarchy of the project is based on the naming convention of the trolley, where the detectors are installed. Each trolley has six sections and each section accommodates one detector. Currently three CMS RPCs trolleys are installed in the GIF++. Trolley 1 (RPC Consolidation) is equipped with spare RPCs, trolley 2 with small glass RPCs and trolley 3 with prototypes of improved RPCs. Detailed information of the trolleys are given here \cite{salvador}. 
A schematic overview of the DCS project is shown in figure \ref{DCS_sys}. 
\bigbreak \bigbreak \bigbreak \bigbreak

\subsection{High and Low Voltage System}
The high and low voltage system is controlled and monitored through the CAEN OPC server. Each gap of a chamber is independently connected to a single high voltage channel which improves the granularity of control. The RPC front-end electronics requires digital and analog power supplies \cite{feb}. Each low voltage line has been shared between two front-end-boards (FEBs) for digital as well as for analog. The low voltage boards are installed in the CAEN main frame and controlled by DCS. 
 
\subsection{Environmental and Gas Parameters}
The performance of RPCs strongly depends on the temperature and pressure of the environment. Hence, it is important to measure the environmental parameters (temperature, pressure and humidity) at different locations. The applied voltage is corrected for the environmental temperature and pressure in order to include its effects. This procedure is described in detail in \cite{env-rpc}.  
Figure \ref{scan_temp}a gives a plot for environmental temperature, pressure and humidity. The environmental and gas sensors (temperature, pressure and humidity) are readable through ADC (analog-to-digital converter) board which is installed in the EASY crate. The JCOP framework gives the opportunity to convert online the ADC counts into physical values. The trending feature provides a comparison among different sensors located at different positions.


\subsection{High Voltage Scan and Stability Test}
\begin{figure}[htp]
\centering
\begin{tabular}{cc}
\hspace{-0.5cm}
\includegraphics[scale=0.32,trim=50 70 20 90,clip]{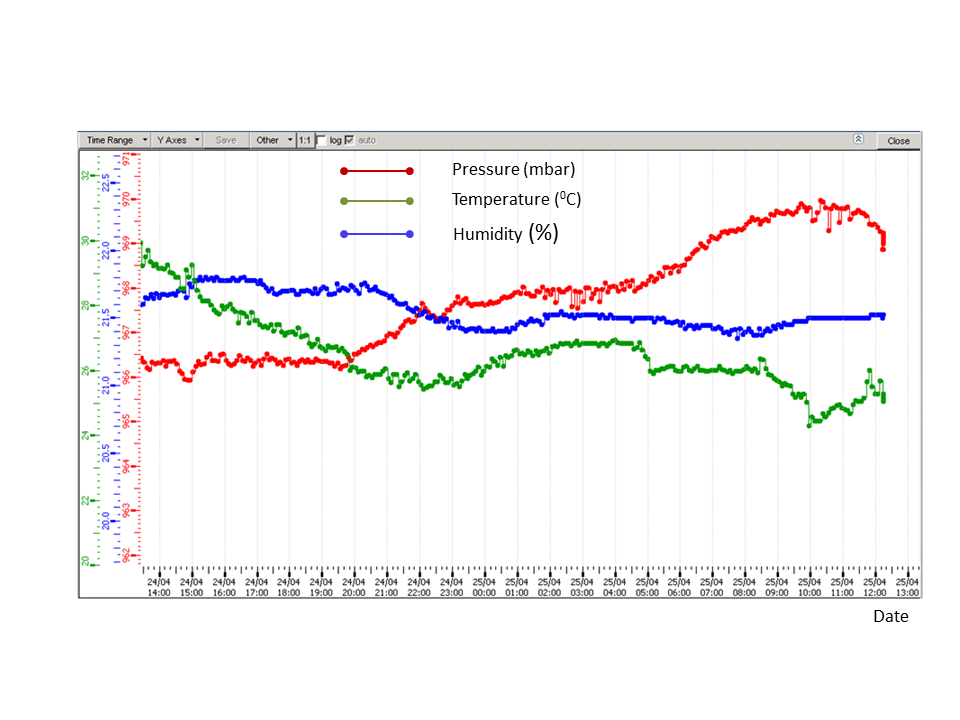}
& \hspace{-0.60cm} \includegraphics[scale=0.334,trim=50 100 45 80,clip]{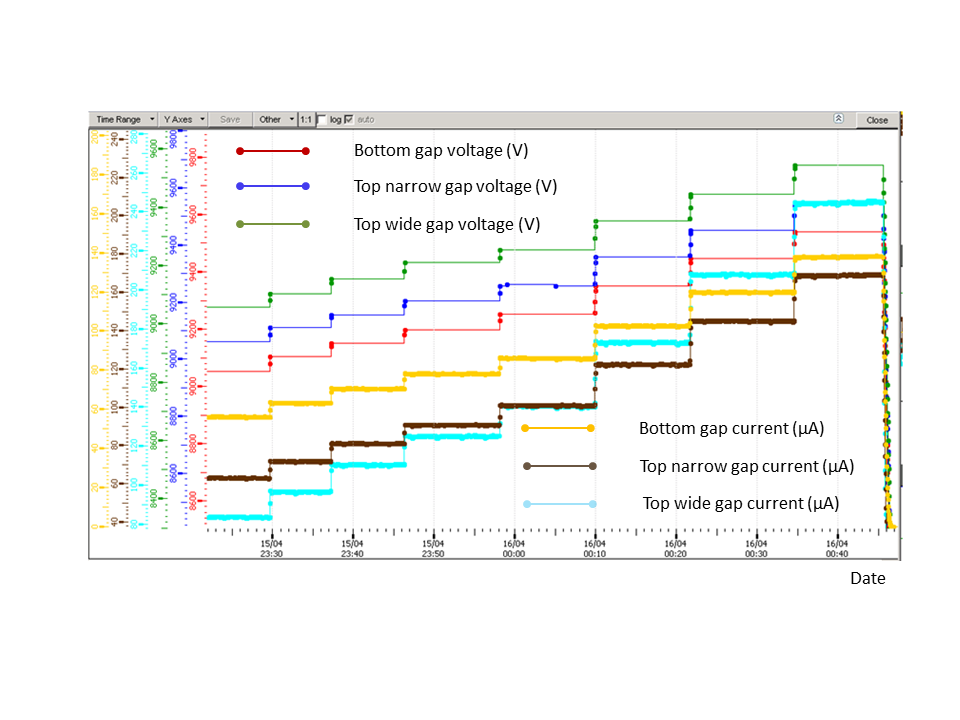}\\
   ($\mathbf{a}$)\qquad&($\mathbf{b}$)\qquad\\
\end{tabular}
\caption{(a): Environmental temperature ($^o$C), pressure (mb) and humidity (\%). Time is on x-axis while red, blue and green lines show the values of pressure, temperature and humidity respectively on y-axis. Pressure and temperature values used for operating voltage correction of RPCs. (b): A high voltage scan for CMS RE3 chamber. The x-axis shows time while the y-axis shows the voltage (V) and current ($\mu$A) values. The red, blue and green lines are voltages while cyan, brown and orange lines are the corresponding current values for bottom, top narrow and top wide gaps respectively.}\label{scan_temp}
\end{figure}

The project has been designed for R\&D of detectors, hence, should be able to perform high voltage scanning or stability tests. For high voltage scanning, a separate branch has been incorporated in the Finite State Machine (FSM) tree, where the user operates each detector independently. The stability test runs for long time. Based on the requirements, a dedicated manager is used to apply the stability script and restart it automatically. A high voltage scanning plot for one of the CMS RPC chambers (RE3) at GIF++ is shown in figure \ref{scan_temp}b. 

\subsection{FSM and GUI} 
\begin{figure}[htp]
\centering
\includegraphics[scale=0.61,trim=0 175 10 80,clip]{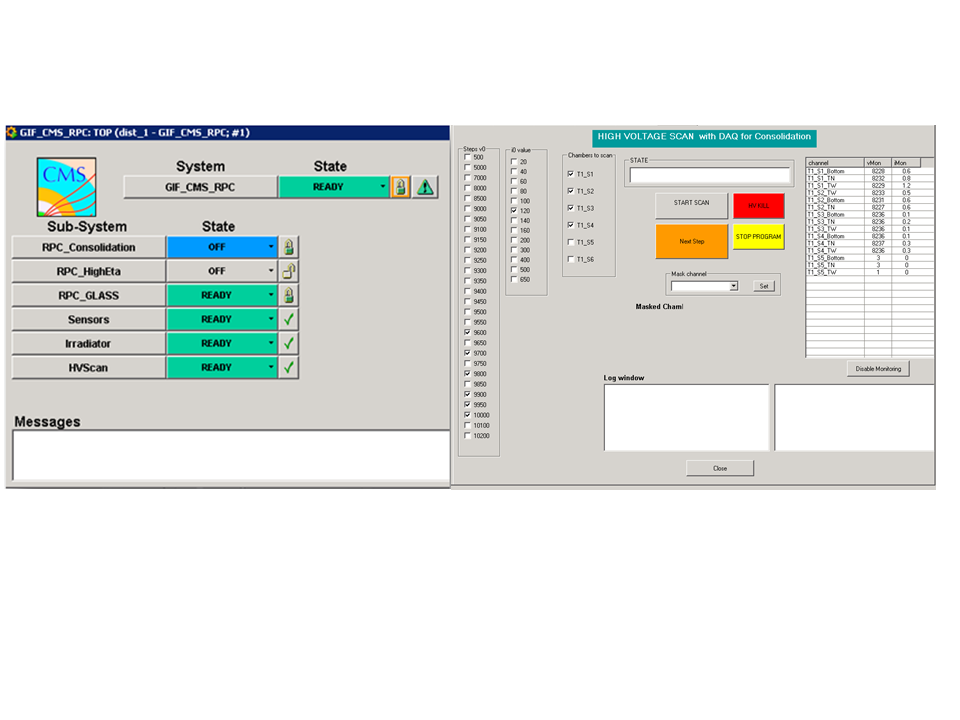}\\
 \caption{FSM main tree and high voltage scan panel using GUI.}
\label{gui}
\end{figure}

The JCOP framework provides FSM toolkits in WinCC-OA based on State Machine Interface (SMI++). It offers an easy, robust and safe way to control the full detector through the definition of a finite number of states, transitions and actions (ON, OFF, STANDBY, Ramping Up, Ramping Down). All the DCS hierarchy nodes are implemented through the FSM mechanism, shown in figure \ref{gui}.\\
WinCC-OA provides a user friendly Graphic User Interface (GUI) panel- an intuitive tool to control, monitor and operate the detectors in safe mode. It provides flexibility to combine text, graphical objects and synoptic diagrams. GUI panels can be used to see the online behavior of the detector in the form of plots, tables and histograms. 

\subsection{DataBase}
To study the behavior of the detector over time and to make offline data analysis, it is necessary to store all the important parameters in a database. WinCC-OA's internal SQL database is used in this project. The stored data can be extracted for the offline analysis using a GUI. 

\section{Detector performance study}
\subsection{RPCs setup at GIF++ and Method for efficiency study}\label{eff_calc}
Four RPC chambers (two RE2 and two RE4 \cite{feb}) are placed parallel to each other in vertical position. Several high voltage scans were performed using different radiation levels (starting from the absence of radiation source), to define the optimal operating voltage of each chamber, called working point (WP). The efficiency (E) for different background radiation levels is calculated using the formula:\\ 
\centerline {E = N$_{RPC}$/N$_{TRACK}$.}
The muon track is reconstructed using three RPC planes and extrapolated to the RPC under test, looking to the closest cluster (a strip or set of continuous strips). N$_{TRACK}$ corresponds to the number of muons passing through the reference three RPC detectors at the same time. N$_{RPC}$ corresponds to the number of fired clusters in the chamber under test. The dependency of the efficiency E with respect to the effective high voltage HV$_{eff}$ \cite{hv-eff} can be fitted using the sigmoidal curve described by the subsequent formula:\\
\centerline {$\langle$E$\rangle$ = E$_{max}$ / (1 + exp (-$\lambda$ (HV$_{eff}$ - HV$_{50\%}$))}
where E$_{max}$ is the maximum efficiency reached by chambers at HV$\rightarrow \infty$, $\lambda$ is proportional to the slope of sigmoid at flex point and HV$_{50\%}$ is the high voltage at which a chamber reaches 50\% of its maximum efficiency. Working point of a chamber is defined by the formula HV$_{wp}$ = HV$_{knee}$ + 150V, where HV$_{knee}$ is the voltage at which efficiency is 95\% of the maximal one.   

\subsection{Efficiency Results}
\begin{figure}[htp]
\centering
\begin{tabular}{cc}
\hspace{-0.5cm}
\includegraphics[scale=0.30,trim=0 13 0 0,clip]{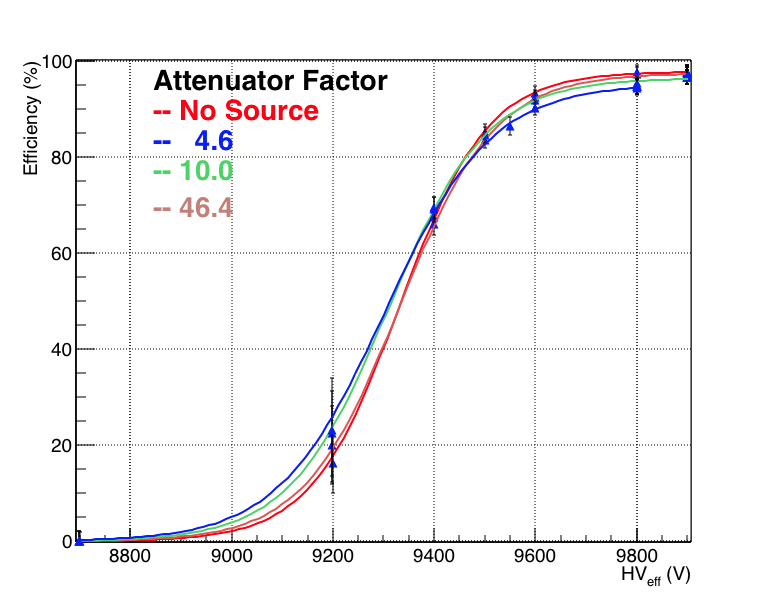}
& \hspace{-0.60cm} \includegraphics[scale=0.329,trim=0 0 0 0,clip]{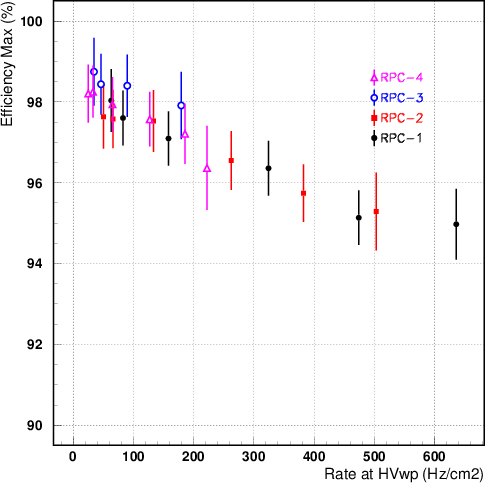}\\
   ($\mathbf{a}$)\qquad&($\mathbf{b}$)\qquad\\
\end{tabular}
\caption{(a): Efficiency vs HV$_{eff}$ for different gamma attenuator factors. (b): Maximum efficiency vs gamma rate at HV$_{wp}$ for four RPCs.}\label{eff}
\end{figure}

In figure \ref{eff}a, the efficiency as a function of HV$_{eff}$ for different radiation levels is shown. The maximum efficiency decreases with the amount of radiation received by the detector. In figure \ref{eff}b, the maximum efficiency as a function of the gamma hits rate is presented for four RPC chambers. The RPCs were placed parallel to each other, RPC-1 being the closest to the source and RPC-4 being the most distant. The radiation doze depends on the distance between the RPCs and the gamma source, RPCs 3 and 4 received a smaller dose as compared to RPCs 1 and 2. 

\section{Conclusion} 
A DCS project for CMS RPCs has successfully been implemented and tested in the CERN GIF++. Since June, 2015 the project is running in a stable state, operating the detectors and archiving the data. The hardware integrated in the project, fully controls high voltage scanning and stability tests. The environmental and gas sensors are included and used for Temperature/Pressure corrections. Gas flow-meters are read through central DCS at GIF++ and the data are used to study the behavior of different gases. All useful parameters are archived in the internal database for offline analysis. As the project is designed for detector R\&D studies, any new hardware can be added easily and safely.\\
The performance of the CMS RPC chambers at GIF++ has been studied and compared at different radiation levels. At a rate of 600 Hz-cm$^{-2}$, the Eff$_{max}$ of the chamber was 95\%. 
\acknowledgments We wish to congratulate our colleagues in the CERN Engineering- (EN) and Physics- Department (PH) for successful operation of the GIF++. We thank the technical and administrative staff at CERN, other CMS institutes and RPC group. Many thanks to ATLAS colleague Marino Romano for his technical support to develop the project.   


\end{document}